\documentclass[10pt, conference]{IEEEtran}
\IEEEoverridecommandlockouts
\pdfoutput=1
\usepackage{cite}
\usepackage{amsmath,amssymb,amsfonts}
\usepackage{physics}
\usepackage{algorithmic}
\usepackage{graphicx}
\usepackage{textcomp}
\usepackage{xcolor}
\usepackage{tabularx}
\def\BibTeX{{\rm B\kern-.05em{\sc i\kern-.025em b}\kern-.08em
    T\kern-.1667em\lower.7ex\hbox{E}\kern-.125emX}}
\usepackage{hyperref}
\hypersetup{
    colorlinks=true,
    allcolors=magenta,
}
\usepackage[utf8]{inputenc}
\usepackage[T1]{fontenc}
\usepackage{qcircuit}

\DeclareUnicodeCharacter{2009}{ }
\DeclareUnicodeCharacter{03B2}{\ensuremath{\beta}}

\begin{document}

\title{Averting multi-qubit burst errors\\in surface code magic state factories\\

\thanks{
This work is funded in part by EPiQC, an NSF Expedition in Computing, under award CCF-1730449; in part by STAQ under award NSF Phy-1818914/232580; in part by the US Department of Energy Office of Advanced Scientific Computing Research, Accelerated Research for Quantum Computing Program; and in part by the NSF Quantum Leap Challenge Institute for Hybrid Quantum Architectures and Networks (NSF Award 2016136), in part based upon work supported by the U.S. Department of Energy, Office of Science, National Quantum Information Science Research Centers, and in part by the Army Research Office under Grant Number W911NF-23-1-0077. The views and conclusions contained in this document are those of the authors and should not be interpreted as representing the official policies, either expressed or implied, of the U.S. Government. The U.S. Government is authorized to reproduce and distribute reprints for Government purposes notwithstanding any copyright notation herein.

F.C. is the Chief Scientist for Quantum Software at Infleqtion and an advisor to Quantum Circuits, Inc.

$^\dagger$\href{mailto:jchadwick@uchicago.edu}{jchadwick@uchicago.edu}.
}
}

\makeatletter 
\newcommand{\linebreakand}{%
  \end{@IEEEauthorhalign}
  \hfill\mbox{}\par
  \mbox{}\hfill\begin{@IEEEauthorhalign}
}
\makeatother 




\author{\IEEEauthorblockN{
Jason D. Chadwick$^\dagger$,
Christopher Kang,
Joshua Viszlai,
Sophia Fuhui Lin,
and Frederic T. Chong
}
\IEEEauthorblockA{\textit{Department of Computer Science, University of Chicago}
}
}

\maketitle
\thispagestyle{plain}
\pagestyle{plain}

\begin{abstract}
Fault-tolerant quantum computation relies on the assumption of time-invariant, sufficiently low physical error rates. However, current superconducting quantum computers suffer from frequent disruptive noise events, including cosmic ray impacts and shifting two-level system defects.  Several methods have been proposed to mitigate these issues in software, but they add large overheads in terms of physical qubit count, as it is difficult to preserve logical information through burst error events. 

We focus on mitigating multi-qubit burst errors in magic state factories, which are expected to comprise up to 95\% of the space cost of future quantum programs. Our key insight is that magic state factories do not need to preserve logical information over time; once we detect an increase in local physical error rates, we can simply turn off parts of the factory that are affected, re-map the factory to the new chip geometry, and continue operating. This is much more efficient than previous more general methods, and is resilient even under many simultaneous impact events. Using precise physical noise models, we show an efficient ray detection method and evaluate our strategy in different noise regimes. Compared to existing baselines, we find reductions in ray-induced overheads by several orders of magnitude, reducing total qubitcycle cost by geomean $6.5\times$ to $13.9\times$ depending on the noise model. This work reduces the burden on hardware by providing low-overhead software mitigation of these errors.
\end{abstract}


\section{Introduction}\label{sec:intro}

\begin{figure*}[htp]
    \centering
    \includegraphics[width=\textwidth]{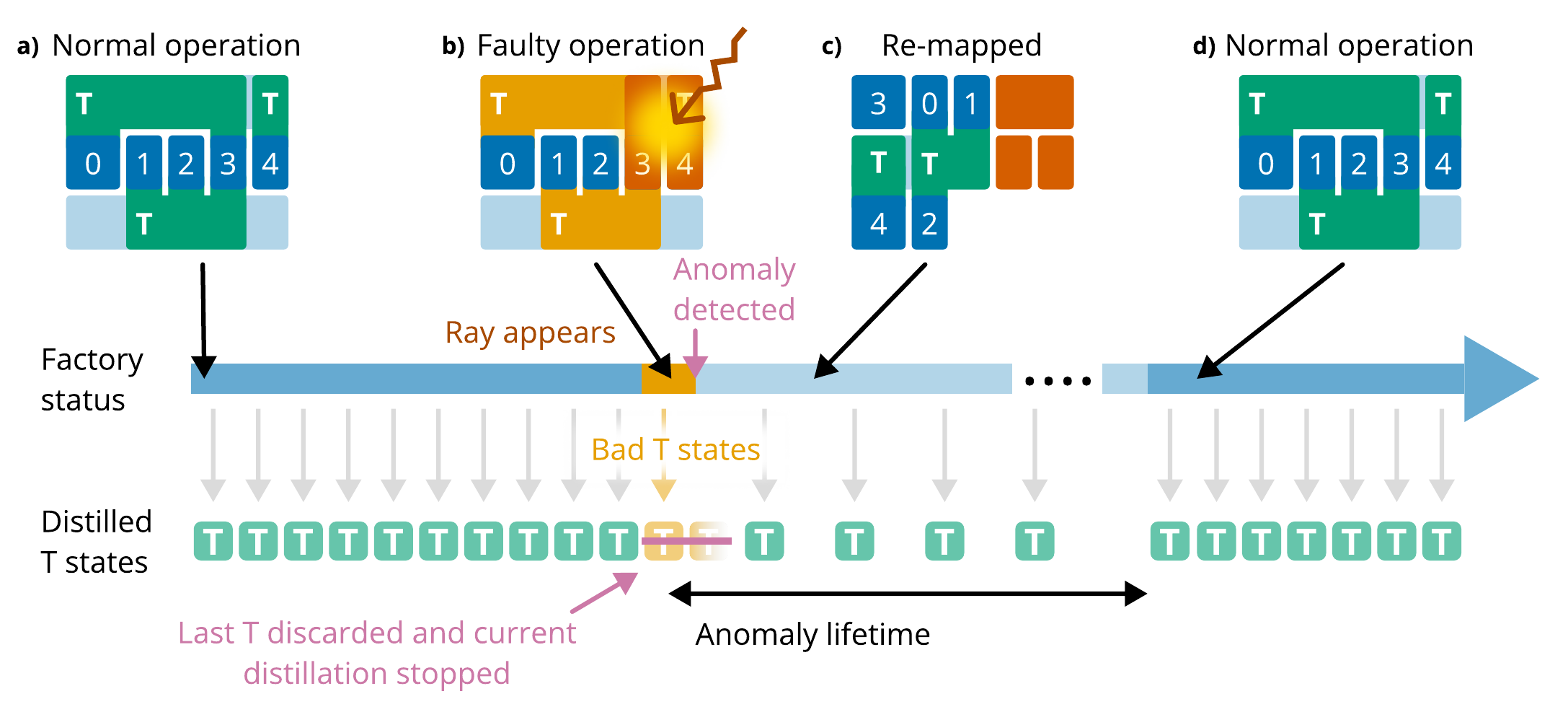}
    \caption{Overview of our method, showing the timeline of a magic state factory before and after a cosmic ray event. (a) The factory is operating normally. (b) A cosmic ray hits the chip, severely affecting physical qubit error rates nearby and causing the factory to output low-quality magic states. (c) The cosmic ray is detected after some delay. The distilled magic states in the buffer (yellow) are discarded, the affected physical qubits are turned offline, and the factory is re-mapped to avoid using the offline areas. The factory can continue to operate at reduced speed. (d) The affected physical qubits recover from the ray impact and are turned back online. The factory resumes normal operation.}
    \label{fig:hero}
\end{figure*}


Noise is the perpetual adversary of practical quantum computing. While recent advances have continued to yield improved physical qubits \cite{stehlik_tunable_2021, philips_universal_2022, weinstein_universal_2023, evered_high-fidelity_2023, moses_race-track_2023, klimov_optimizing_2024}, it is generally accepted that physical qubits will never reach the error rates of \textless$10^{-10}$ required for practical applications \cite{beverland_assessing_2022}. Thus, quantum error correction (QEC) is critical to build and scale fault-tolerant quantum computing. QEC suppresses errors by encoding logical qubits into many physical qubits. Recent experiments have shown primitive versions of QEC \cite{egan_fault-tolerant_2021, postler_demonstration_2022, takeda_quantum_2022, google_quantum_ai_suppressing_2023, weinstein_universal_2023, bluvstein_logical_2024} and support the aim to achieve \textless$10^{-10}$ logical error rates.


Most theoretical QEC research inacurately assumes constant physical qubit error rates across both space (qubit position) and time. However, on superconducting platforms,  gate error rates, readout error rates, and coherence times can vary by orders of magnitude across physical qubits \cite{google_quantum_ai_suppressing_2023, kim_evidence_2023} and experience significant drifts in error rates on the timescale of hours or even minutes \cite{bialczak_1_2007, proctor_detecting_2020}; these nonuniform errors pose significant problems for future fault-tolerant quantum computing \cite{hanks_decoding_2020, clader_impact_2021, berke_transmon_2022, iolius_performance_2022, carroll_subsystem_2024}. To alleviate this problem, hardware providers typically calibrate their devices at least once a day \cite{klimov_optimizing_2024}.

Time-varying noise can be especially damaging: superconducting hardware suffers from 
cosmic ray impacts \cite{orrell_sensor-assisted_2021, wilen_correlated_2021, mcewen_resolving_2022, li_direct_2024}, which can drastically reduce the coherence of 20+ qubits at a time and occur as often as once every 10 seconds on a 27-qubit chip \cite{mcewen_resolving_2022}, and shifting two-level system (TLS) defects \cite{meisner_probing_2018, klimov_fluctuations_2018, schlor_correlating_2019, muller_towards_2019, carroll_dynamics_2022}, which are long-lived, localized reductions in qubit coherence that can instantaneously jump in intensity \cite{klimov_fluctuations_2018}. These events are unpredictable and cannot be mitigated passively via periodic calibration.  

Previous work has established several methods to actively mitigate cosmic rays and TLSs through software. These techniques include code concatenation \cite{xu_distributed_2022}, quickly moving logical qubits away from degraded regions \cite{sane_fight_2023}, modifying stabilizers to ignore the affected region \cite{siegel_adaptive_2023, strikis_quantum_2023}, and temporarily expanding code distance and re-executing the decoder \cite{suzuki_q3de_2022}. However, these methods all incur significant qubit overhead ($\sim$$2$-$10\times$ or more) to maintain the baseline logical error rate, as they are designed to preserve information in logical qubits even during a cosmic ray event.

In this work, we focus on mitigating bursts of error in magic state factories. Distillation is projected to have a much larger footprint than other program components in most fault-tolerant programs, potentially consuming up to 95\% of the program's spacetime cost \cite{beverland_assessing_2022, babbush_encoding_2018, blunt_compilation_2024}. 

A key feature of these factories is that they do not store program  data, so we do not need to spend valuable quantum resources to preserve their integrity amid a disruptive error event. Figure \ref{fig:hero} shows the main idea of our work: once we detect an increase in local physical error rates, we can simply turn off parts of the factory that are affected, re-map the factory to the new chip geometry, and continue operating. This is much more efficient than previous general methods that attempt to preserve the information encoded within logical qubits.



To test our method, we design detailed physical noise models that reproduce two key effects of cosmic ray impacts on qubits: directly reducing qubit coherence times and indirectly scrambling TLS defects. We design methods to detect these events by counting windowed stabilizer syndrome measurement results, and then evaluate the realistic overhead of mitigating cosmic rays using several baseline methods and our own. We find considerable reductions in total qubitcycle cost per distillation using our method, effectively reducing the burden on hardware designers by providing low-overhead software mitigation of multi-qubit correlated burst errors.

This paper is structured as follows: In Section \ref{sec:background}, we introduce essential background information on device noise and magic state distillation. In Section \ref{sec:noise-models}, we describe the noise models we use in our evaluation. In Section \ref{sec:detecting}, we describe how to rapidly and efficiently detect a disruptive noise event. In Section \ref{sec:remapping}, we show how we adapt a magic state factory to fit into a deficient (partially-offline) footprint. In Section \ref{sec:evaluation}, we introduce baselines and describe evaluation methodology. In Section \ref{sec:results}, we examine the performance of our re-mapping method in various noise regimes and compare to other existing methods. Finally, in Section \ref{sec:discussion}, we discuss the broader implications of our work.

\section{Background}\label{sec:background}

\subsection{Two-level systems and cosmic rays}\label{sec:background/rays}

Cosmic rays can affect qubits in several ways. In the following section, we describe the physical processes that cause multi-qubit burst errors upon a cosmic ray impact.

A two-level system (TLS) defect is an unwanted resonance on a superconducting chip \cite{meisner_probing_2018, schlor_correlating_2019}. If a qubit's transition frequency is too close to the frequency of a TLS, the qubit will interact with the TLS and suffer a strong decrease in coherence \cite{klimov_fluctuations_2018}. Even small superconducting chips typically have many localized TLSs \cite{klimov_fluctuations_2018, muller_towards_2019, carroll_dynamics_2022}. On architectures with tunable-frequency qubits, TLS defects can be avoided by operating qubits in frequency regimes that are not affected \cite{klimov_optimizing_2024}. However, as TLS resonant frequencies can abruptly shift during a computation \cite{klimov_fluctuations_2018}, a static calibration would not ensure stability during long computations. A dynamic retuning strategy \cite{kelly_scalable_2016} could be employed to ensure stability during long surface code computations, but there is no guarantee that this would be sufficient, as calibration is a highly complex problem that cannot be solved by local optimization \cite{klimov_optimizing_2024}.

A cosmic ray impact on a superconducting qubit chip causes a burst of quasiparticles that reduce qubit coherence in an area around the impact site \cite{orrell_sensor-assisted_2021, wilen_correlated_2021, mcewen_resolving_2022, li_direct_2024, cardani_disentangling_2023}.  Cosmic rays are well-understood but still difficult to mitigate: gap engineering \cite{sun_measurements_2012, pan_engineering_2022, marchegiani_quasiparticles_2022, kamenov_suppression_2023, mcewen_resisting_2024} can reduce the strength of the effect, but  these approaches require different fabrication methods or materials, posing an additional challenge to building large-scale high-performance chips.

Furthermore, other detrimental effects may still persist in spite of gap engineering. Most relevant to our work is an effect known as \emph{TLS scrambling} \cite{thorbeck_two-level-system_2023}, where TLS defects near the ray impact location can experience large shifts, leading to random, persistent increases in qubit error rates. These shifted TLSs may remain in their new configurations for hours \cite{klimov_fluctuations_2018}; thus, qubits must be re-tuned to shift away from the defects.

Alternatively, processors can be protected from external radiation with metallic shielding \cite{iaia_phonon_2022, pan_engineering_2022}. However, shielding complicates device construction and does not guarantee invulernability from all radiation, as harmful radiation can still come from sources within device materials themselves \cite{cardani_disentangling_2023}. Even a single burst error event in a program can be catastrophic; thus, it is still worth exploring low-overhead methods that can tolerate these events. 

Cosmic ray impacts thus pose significant problems for scaling quantum hardware into the fault-tolerant era, both through direct interaction with qubits and through TLS scrambling. In this work, we aim to provide a software solution to reduce the impact of these disruptive noise processes on magic state factories, reducing the burden on hardware designers.

\subsection{Error correction and the surface code}

The surface code \cite{kitaev_quantum_1997, kitaev_fault-tolerant_2003, bravyi_universal_2005} is a promising approach towards quantum error correction due to its planar topology and relatively high threshold. Each logical qubit is encoded in a rectangular grid of physical qubits (Figure \ref{fig:background}b). The surface code can fault-tolerantly perform quantum operations in the Clifford group (such as X, Z, H, S, and CNOT) using lattice surgery \cite{horsman_surface_2012, litinski_game_2019, fowler_low_2019, chamberland_universal_2022}, which involves creating, growing, merging, and measuring patches of surface code (example in Figure \ref{fig:background}d). However, the Clifford group alone is not sufficient to express arbitrary quantum operations; for a universal gate set, the typical choice is to include one additional operation known as the T gate ($\text{T} = \sqrt{\text{S}} = \sqrt[4]{\text{Z}}$), which allows for universal quantum computation \cite{chamberland_universal_2022}. 

This T gate cannot be natively implemented in a fault-tolerant way on the surface code. As shown in Figure \ref{fig:background}a, a T operation can be performed using a so-called \emph{magic state} $\ket{\text{T}} = \frac{1}{\sqrt{2}}(\ket 0 + e^{i \pi/4}\ket 1)$. To achieve sufficiently high fidelities, these logical magic states must be prepared in a special (and expensive) process known as \emph{magic state distillation}.

\subsection{Magic state distillation in the surface code}


\begin{figure}
    \centering
    \includegraphics[width=\linewidth]{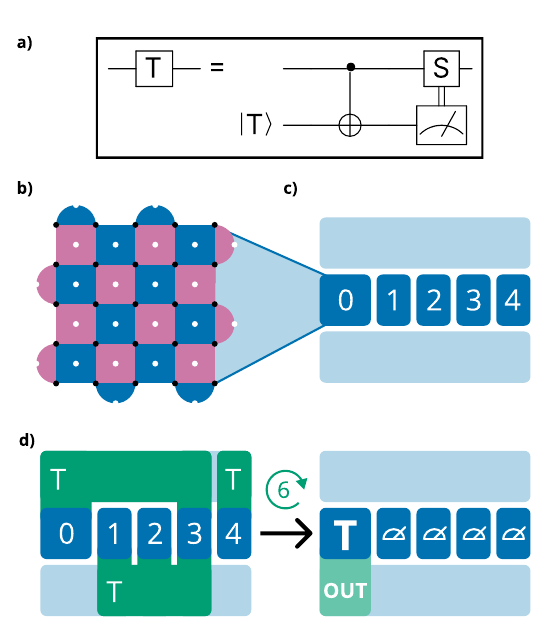}
    \caption{Background on magic state distillation. (a) A high-fidelity magic state can be consumed in a fault-tolerant circuit to perform a logical T gate. (b) A surface code patches is a rectangular grid of physical qubits that can act as one logical qubit. (c) The magic state factory layout from \cite{litinski_magic_2019}, made up of five logical qubits (dark blue) and two routing space channels (light blue). (d) One step of a distillation. In this step, three faulty T rotations are applied to the qubits. 15-to-1 distillation uses 15 faulty T states in 6 rounds of operations, and outputs one higher-fidelity T state at the end.}
    \label{fig:background}
\end{figure}

Magic state distillation \cite{bravyi_magic-state_2012, litinski_magic_2019} is a process which converts noisy \emph{physical} T states to high-fidelity \emph{logical} T states. A physical qubit in the T state (created by initializing the qubit to $\ket 0$ and then applying physical H and T gates) is first \emph{injected} into the surface code via code expansion \cite{li_direct_2024}, creating a noisy logical T state. The fidelity of this logical T state cannot be greater than the fidelity of the initial physical T state; distillation is the process of combining several copies of noisy logical T states in a circuit to generate a smaller number of high-quality \emph{output} T states. In \emph{15-to-1 distillation} \cite{bravyi_magic-state_2012, litinski_magic_2019}, 15 of these noisy logical T states are used to create one higher-fidelity logical T state. If the 15 input states each have infidelity $p$, the output state of 15-to-1 distillation will have infidelity $\mathcal{O}(35p^3)$ (assuming perfect surface code operations during distillation). Ref. \cite{litinski_magic_2019} provides an efficient distillation circuit that uses five logical qubits to consume the 15 noisy T states in six rounds of lattice surgery operations. The layout of this factory is shown in Figure \ref{fig:background}c.

\section{Noise models}\label{sec:noise-models}

\begin{figure*}
    \centering
    \includegraphics[width=\textwidth]{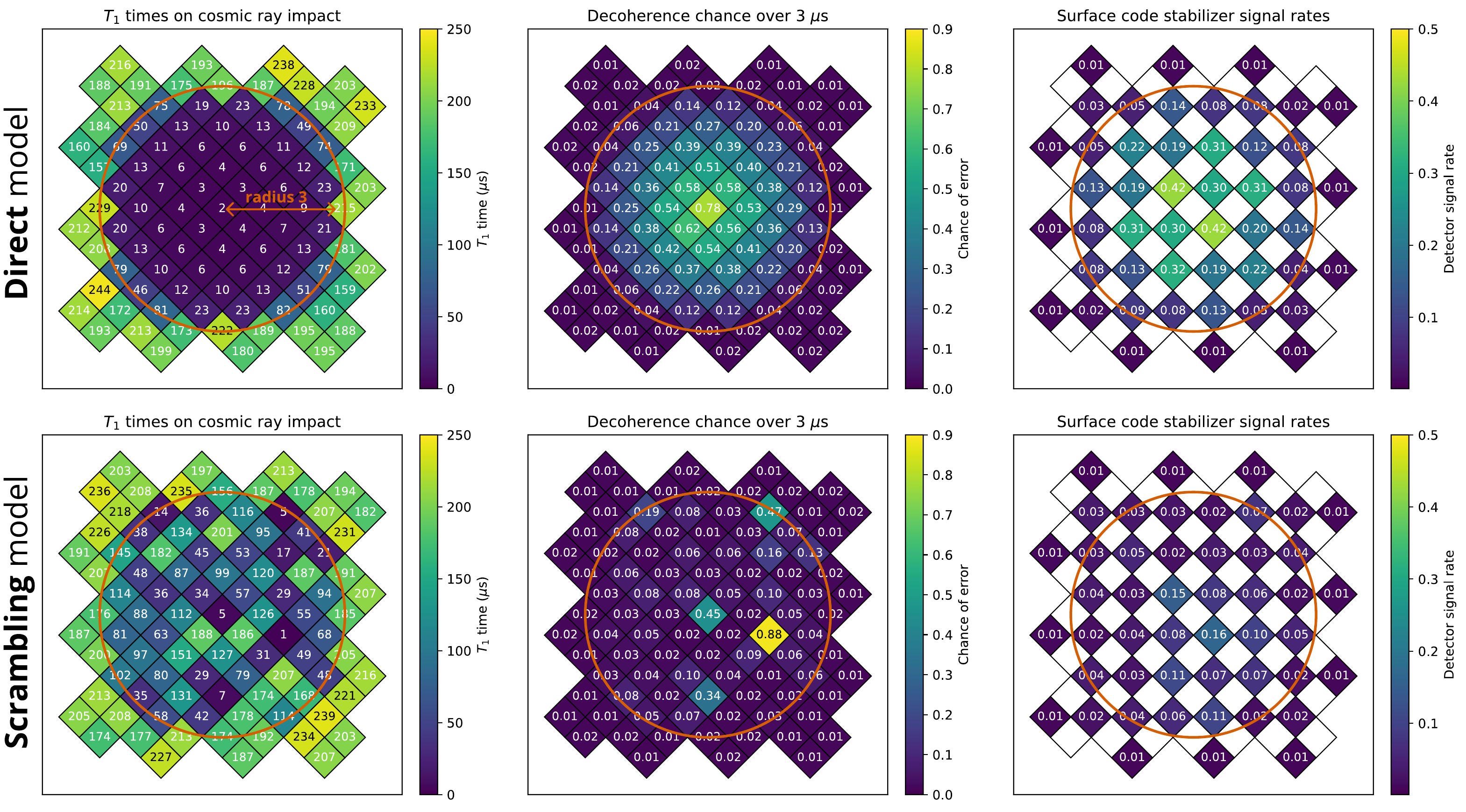}
    \caption{The two cosmic ray noise models studied in this work. From left to right, we show the effect of a representative ray on the physical qubit $T_1$ times, the decoherence error rate over $3\mu$s (the time period used in \cite{mcewen_resolving_2022}), and the chance that each surface code stabilizer measurement will detect an error. \emph{Top:} The \textbf{Direct} model is based on \cite{mcewen_resolving_2022}. A ray impact directly affects the $T_1$ times of qubits within some radius, becoming less severe with distance. \emph{Bottom:} The \textbf{Scrambling} model is based on \cite{wilen_correlated_2021, thorbeck_two-level-system_2023}. A ray impact scrambles TLS defects in some area, leading to unpredictable and long-lasting effects on qubit coherence.}
    \label{fig:ray-on-patch}
\end{figure*}

We use a circuit-level noise model in which each physical qubit has its own set of parameters; we then build our cosmic ray models on this foundation.

\subsection{Physical qubit noise model}

We simulate physical qubit noise in Stim \cite{gidney_stim_2021}. Each qubit is assigned values for $T_1$, $T_2$, single-qubit gate error rate $p_1$, and measure/reset error rate $p_{\text{MR}}$. Each qubit pair is assigned a two-qubit gate error rate $p_2$. In this work, we assign default values of these parameters based on the Sycamore processor benchmarked in \cite{google_quantum_ai_suppressing_2023}, with mean (standard deviation) set to $T_1$: $20(2) \mu \text{s}$ and $T_2$: $30(5) \mu \text{s}$. We apply idling errors during gate operations and add small additional gate errors to give total error mean (standard deviation) of $p_1^\text{tot}$: $0.0008(5)$, $p_2^\text{tot}$: $0.005(3)$, $p_{\text{MR}}^\text{tot}$: $0.020(5)$, again modeling the Sycamore processor \cite{google_quantum_ai_suppressing_2023}. We then improve all the above values by $10\times$ to model future device performance and ensure we are comfortably below the fault-tolerant threshold. $T_1$ and $T_2$ values for each qubit are sampled from a normal distribution $\mathcal N(\mu, \sigma)$, while error rates $p_1, p_2, p_{\text{MR}}$ are sampled from a lognormal distribution with equivalent mean and standard deviation.

\subsection{Modeling cosmic rays}\label{sec:evaluation/noise}

We employ two distinct cosmic ray models:

\subsubsection{Direct interaction models}

As explained in Section \ref{sec:background/rays}, a cosmic ray impact can cause a large drop in qubit $T_1$ times for all qubits within some radius of the impact location. Reference \cite{mcewen_resolving_2022} measured the chance of qubit idle errors during 3 $\mu$s time periods, observing that a ray causes an extremely high rate of errors which linearly decreases with distance from the center (note that this does not mean that the $T_1$ times decrease linearly). We recreate this same relationship in our \textbf{Direct} noise model, as shown in Figure \ref{fig:ray-on-patch}. Within a given radius $r_{\text{CRE}}$, qubit $T_1$ times range from $f_{T_1}T_1^{\text{init}}$ at the center to $T_1^{\text{init}}$ at the edge, where $T_1^{\text{init}}$ is the default coherence time of a qubit without a ray. The parameter $f_{T_1}$ is the $T_1$ reduction factor at the center of the ray, where it is strongest. We test three values of $f_{T_1}$: $(0.1, 0.01, 0.001)$.

We assume that ray impacts follow a Poisson distribution with rate $\Gamma$ (events / qubit / second). Ref. \cite{mcewen_resolving_2022} observed a rate of 1 impact per 10 seconds on their chip of 27 qubits. Qubits recover from a ray impact in time $T_{\text{offline}}$. Ref. \cite{mcewen_resolving_2022} observed that the effect on qubit $T_1$ values decays within around 50 ms. The value $\Gamma \times T_{\text{offline}}$ determines the proportion of time that a region of the device spends in normal operation, recovering from one ray, or recovering from multiple rays. This value is important to determine the overhead of a mitigation method, as it sets requirements on how many simultaneous rays the mitigation method must be able to tolerate, and, in our method, determines how much time the factory will spend in a partially-disabled state.

To summarize, the \textbf{Direct} noise model parameters are:

\begin{itemize}
    \item $r_{\text{CRE}}$ (ray radius): The radius of the ray impact, in terms of qubit spacings on the chip. The effect on qubit $T_1$ time decays with increasing distance from center. We set this parameter between 1 and 4.
    
    \item $f_{T_1}$ ($T_1$ reduction factor at center): The strongest reduction in qubit $T_1$ times at the center of the ray impact. We set this parameter between 0.1 (90\% reduction at center) and 0.001 (99.9\% reduction at center).

    \item $\Gamma$ (ray impact rate per qubit per second) and $T_{\text{offline}}$ (recovery time): These two numbers together determine the proportion of time spent recovering from rays compared to the time spent in normal operation.
\end{itemize}

\subsubsection{TLS scrambling}

Another harmful effect of cosmic rays is their tendency to scramble the resonant frequencies of TLSs around the ray impact site \cite{thorbeck_two-level-system_2023}. As mentioned in \ref{sec:background/rays}, TLS frequency shifts may bring them closer to resonance with the qubits, yielding an unpredictable decrease in $T_1$ time. 

As in the \textbf{Direct} model, we model the ray impact with some radius $r_{\text{CRE}}$. Upon a ray impact, each qubit within the radius has its $T_1$ time scrambled to a uniformly-random value between $0.01T_1^{\text{init}}$ and $T_1^{\text{init}}$ \cite{thorbeck_two-level-system_2023}, as seen in Figure \ref{fig:ray-on-patch}.

While in the \textbf{Direct} model, the ray impact naturally decays on the timescale of tens of milliseconds, the TLS scrambling effect may last longer than the life of the factory, as the frequency of each TLS may not shift again for hours or days \cite{klimov_fluctuations_2018}. Thus, in this model we view $T_{\text{offline}}$ as the ``recalibration'' time, which is the time to find new operating frequencies for the affected qubits such that they are sufficiently far from the new TLS frequencies. Finding operating frequencies for qubits can be a difficult optimization problem; it is unclear how quickly this recovery operation could be performed on-demand, but it may take seconds to minutes with current state-of-the-art calibration routines \cite{klimov_optimizing_2024}.

To summarize, the \textbf{Scrambling} noise model parameters are:

\begin{itemize}
    \item $r_{\text{CRE}}$ (ray radius): The radius of the ray impact, in terms of qubit spacings on the chip. The effect on qubit $T_1$ time decays with increasing distance from center. We set this parameter between 1 and 4.

    \item $\Gamma$ (ray impact rate per qubit per second) and $T_{\text{offline}}$ (recovery time): These two numbers together determine the proportion of time spent recovering from rays, compared to the time spent in normal operation.
\end{itemize}

\section{Detecting burst error events}\label{sec:detecting}


As we discuss in  Sections \ref{sec:evaluation/baselines-require-instant} and \ref{sec:results}, ray detection latency greatly affects mitigation overheads. 
Prior works typically either modeled cosmic ray impacts with highly simplified noise models \cite{suzuki_q3de_2022, siegel_adaptive_2023}, making detection of cosmic ray impacts relatively easy, or have simply not specified a detection method at all \cite{xu_distributed_2022, sane_fight_2023}. 
Under our noise model, we find that some ray impacts are in fact quite difficult to quickly detect, necessitating a method that can tolerate long detection latencies.

We employ a simple and low-cost method to detect a noise anomaly: identify an increase in the rate of erroneous stabilizer syndrome measurements, as suggested in previous work \cite{suzuki_q3de_2022, siegel_adaptive_2023}. The idea is to repeatedly count the number of recent syndromes and signal a detection if the count exceeds some threshold. Additionally, Ref. \cite{siegel_adaptive_2023} proposes using the DBSCAN (density-based spatial clustering of applications with noise) algorithm to cluster the above-threshold stabilizers and filter out noise; however, this step adds significant computational overhead compared to just counting syndrome measurements. We employ an alternative with lower computational cost which similarly boosts our detector's sensitivity. 

\begin{figure}
    \centering
    \includegraphics[width=\linewidth]{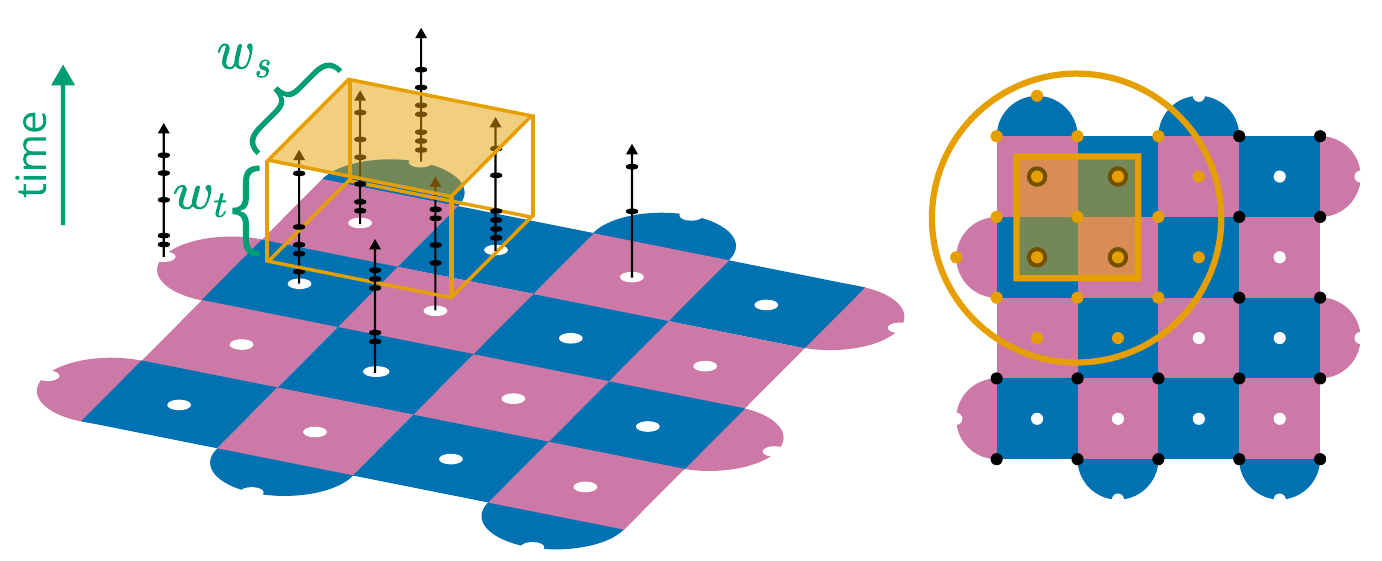}
    \caption{We detect cosmic rays by counting stabilizer syndromes in spatiotemporal windows. \emph{Left:} Each stabilizer produces a syndrome when its measured parity differs from the expected value. We can count the number of syndromes in a $w_s \times w_s \times w_t$ window, and trigger a ray detection event if this count exceeds some threshold. \emph{Right:} When a detection event is triggered, we turn off all physical qubits within the offline flag radius $r_{\text{off}}$.}
    \label{fig:windowing}
\end{figure}

\subsection{Burst error detection via spatial windowing}

We find that we can effectively detect anomalous areas of qubits by counting stabilizer detection events within small overlapping spatiotemporal windows of size $w_s \times w_s \times w_t$ (an example is shown in Figure \ref{fig:windowing}). In this work, we empirically set the spatial window sizes to $w_s = \lceil r_{\text{CRE}}/2 \rceil + 1$ for \textbf{Direct} model and  $\lceil r_{\text{CRE}} \rceil$ for \textbf{Scrambling} model, and set the temporal window size $w_t = c_T = 6d_m$ cycles, where $d_m$ is the temporal distance of the surface code used in the factory. For each window $i$, we calculate the default rate of nontrivial syndrome measurements $p_{\text{syn}, i}$. We assume the distribution of windowed syndrome counts to be binomial $B(w_s \times w_s \times w_t, p_{\text{syn}, i})$. We then choose a desired false positive rate (FPR) and set a threshold $n_{\text{th}}$ such that sampling from the baseline distribution $B(w_s \times w_s \times w_t, p_{\text{syn}, i})$ will yield a value exceeding $n_{\text{th}}$ with probability below the specified FPR.

Each surface code cycle, for each window, we sum the past $(w_s \times w_s \times w_t)$ stabilizer detections and compare to $n_{\text{th}}$. If the sum is greater than the threshold, the window triggers a ray detection, and all qubits within the window are turned off. The spatial windowing approach approximates the DBSCAN method suggested in \cite{siegel_adaptive_2023} with much lower computational overhead. The storage and computational overheads of this anomaly detection method are trivial compared to what is required for real-time decoding \cite{beverland_assessing_2022}, which must already be implemented on any surface code quantum computer.

For the noise models that we consider, qubits within the impacted region are not uniformly affected by the ray (see Section \ref{sec:evaluation/noise}). Ideally, we want to be confident that we will detect \emph{all} anomalous qubits within $r_{\text{CRE}}$ and turn them off. However, in our noise models, not all spatial windows within the ray radius have significantly boosted signal rates. To address this problem, when a detection window triggers, we turn off all qubits within an enlarged area around the window defined by the parameter $r_{\text{off}}$, as shown in Figure \ref{fig:windowing}. We empirically set this to $r_{\text{CRE}}$ for \textbf{Direct} model and $1.5r_{\text{CRE}}$ for \textbf{Scrambling} model.

\subsection{Detector performance}

\begin{figure}[tp]
    \centering
    \includegraphics[width=\linewidth]{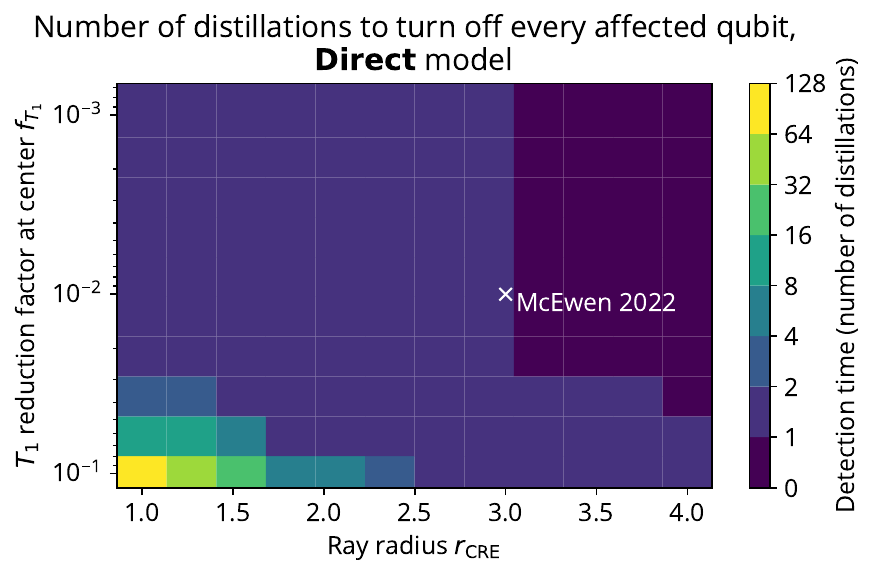}
    \caption{The number of distillations required after a ray impact to flag every physical qubit within \textbf{Direct} ray radius with high probability. Detection becomes signficantly harder as radius and ray severity both decrease. Annotation marks the ray impact rate and approximate radius observed by \cite{mcewen_resolving_2022}.}
    \label{fig:detection-direct}
\end{figure}

\begin{figure}[tp]
    \centering
    \includegraphics[width=\linewidth]{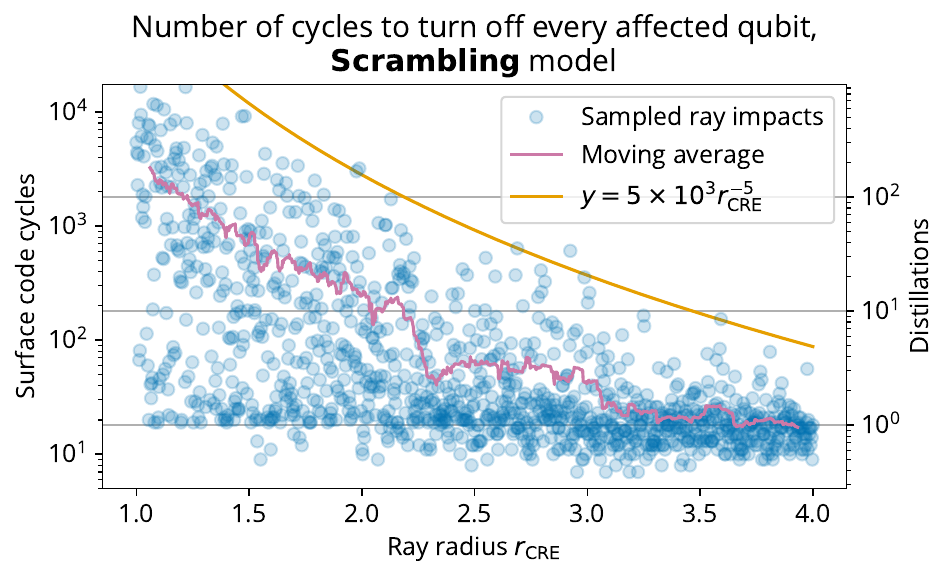}
    \caption{The number of distillations required to flag every physical qubit within \textbf{Scrambling} ray radius with high probability. Detection latency is highly variable for different random ray impacts. We use an approximate upper bound (orange line) for our evaluation.}
    \label{fig:detection-scrambling}
\end{figure}

We now investigate the detection latency $c_D$ for various ray parameters. We set the window detection FPR to $10^{-8}$. For the \textbf{Direct} model, we model a ray impact at the center of a surface code patch (sufficiently large to fully contain the ray) in Stim \cite{gidney_stim_2021} and calculate the number of stabilizer cycles required for \emph{every} physical qubit within $r_{\text{CRE}}$ to be turned off with probability exceeding $1-10^{-6}$. The results are shown in Figure \ref{fig:detection-direct}. We can reliably detect rays in all cases, but the latency of detection grows with smaller or weaker rays. In noise regimes similar to those observed by \cite{mcewen_resolving_2022}, we can fully detect rays within 1-2 distillations.

Similarly to previous work \cite{suzuki_q3de_2022}, we find that \emph{smaller} and less-disruptive cosmic ray impacts lead to overall larger mitigation overheads, due to increased difficulty of detection. Because the effects we consider are significantly less uniform than those considered in previous work, the problem is magnified. If a ray severely alters qubit error rates, the change in syndrome rates is relatively easy to quickly notice, but a smaller shift in error rates (while still affecting the logical error rate significantly) requires a larger temporal window.

Evaluating detector performance in the \textbf{Scrambling} case is more complex, as each ray impact produces a different, random effect. We sample 1200 ray impacts of varying radius. For each, we calculate the resulting stabilizer syndrome signal rates and calculate the number of cycles required to reliably turn off every physical qubit within the radius. The results are shown in Figure \ref{fig:detection-scrambling}. The observed detection latencies are highly variable, ranging over several orders of magnitude. We identify the curve $n_D = 5 \times 10^{3} r_{\text{CRE}}^{-5}$ as the approximate upper bound on the number of distillations required to fully detect the ray.

\section{Re-mapping magic state factories}\label{sec:remapping}

\begin{figure}
    \centering
    \includegraphics[width=\linewidth]{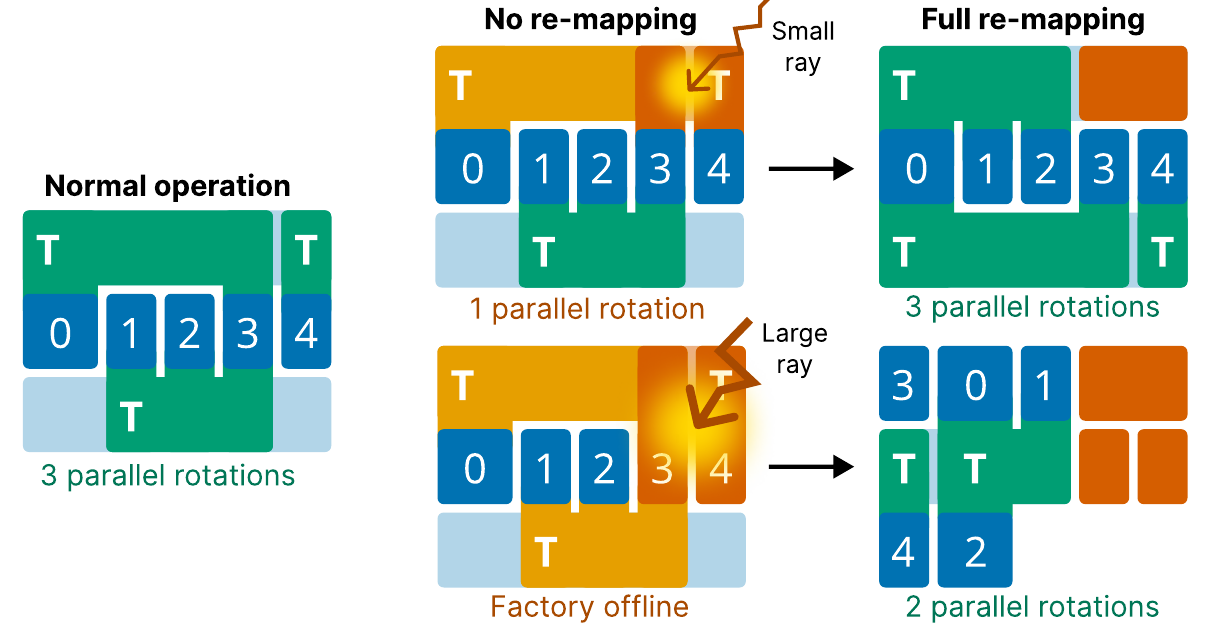}
    \caption{When parts of the chip are forced offline (red patches), re-mapping enables the factory to continue operating in a new configuration, albeit at potentially reduced speed.}
    \label{fig:remapping}
\end{figure}

When a detector window is triggered by a sufficiently-high stabilizer syndrome count, physical qubits lying within the window's flag radius are flagged as faulty. Previous methods attempt to preserve the information in the affected surface code patches by tolerating these bad qubits, either via redundancy or using a large code distance. This is crucial for program qubits, as loss of logical information is disastrous. However, magic state factories do not store long-term logical information, so we can drastically reduce the mitigation cost.

In this section, we explain our burst error mitigation method, which involves maintaining a buffer of distilled T states (set by the ray detection latency) and a strategy to dynamically re-map the factory to avoid defective areas.

\subsection{Distilled T buffer}

We denote $c_T$ as the number of stabilizer cycles required to distill one T state. In the default factory layout (Figure \ref{fig:background}d), $c_T = 6d_m$. Depending on the size and intensity of the anomalous region, it may require more than $c_T$ cycles to confidently detect the ray. Let $c_D$ be the \emph{detection latency} (the number of cycles needed to confidently detect a ray impact under a fixed noise model). When an anomaly is detected, at most $\lceil c_D / c_T \rceil+1$ distilled magic state will have been affected before the detection event, including the state currently being distilled. We thus need to maintain a buffer of $\lceil c_D / c_T \rceil$ magic states, each of which is held in a $d_X \times d_X$ surface code patch. Upon the conclusion of a successful distillation, the oldest buffered state is released to be used in the program and the newly-distilled state takes its place.

\subsection{Re-mapping upon a ray impact}

When some physical qubits within a factory are flagged as faulty, we immediately stop the current distillation and discard all distilled T states in the buffer area (see Figure \ref{fig:hero}). We then attempt to find a valid \emph{re-mapping} of the factory that avoids using the defective areas. Some examples are shown in Figure \ref{fig:remapping}. Re-mapping is done by iterating over all possible placements of the five logical qubits (blue) within the factory footprint, finding the best scheduling of T rotations for each, and picking the layout that can apply all 15 rotations in the fewest number of cycles. If there is no valid layout such that all five logical qubits can be used, the factory is instead turned fully offline until the anomalous region recovers.

During normal operation, a distillation is performed in 6 steps, where each step involves several T rotations in parallel, as shown in the leftmost image of Figure \ref{fig:remapping}. However, once we re-map the factory, we may lose some of this parallelism; in the bottom-right image, only two of the rotations can be performed in parallel. In practice, this means that different ray impacts can have different affects on distillation speed, so we evaluate overhead by simulating many ray impacts and calculating the average cost (and the chance that the factory is forced to turn fully offline) per ray.

\section{Evaluation}\label{sec:evaluation}

\begin{figure}
    \centering
    \includegraphics[width=\linewidth]{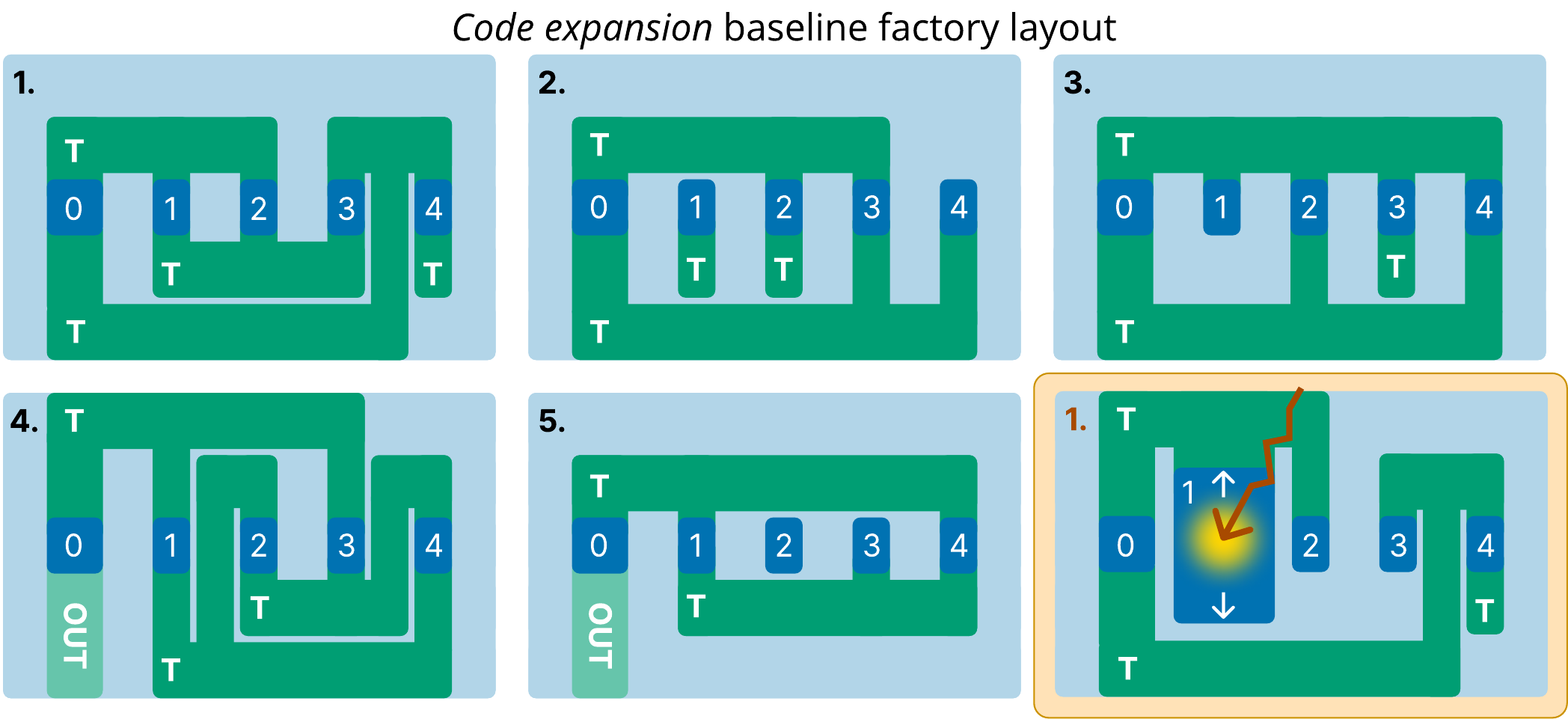}
    \caption{Magic state distillation in the \emph{Code expansion} baseline can be performed in only 5 rounds of operations in the default case, if no anomalies are present. Steps 1-5 show one method of scheduling the 15 T rotations into 5 layers. \emph{Bottom right:} Upon a ray impact, the affected patch can be expanded, allowing it to continue operating. This reverts the distillation to 6 rounds of T rotations.}
    \label{fig:baseline-expanded}
\end{figure}

\subsection{Baselines}\label{sec:evaluation/baselines}

We consider two baseline methods of mitigating cosmic ray impacts: code expansion and concatenated codes. These methods are both designed to protect logical information encoded in the surface code during computation; this satisfies a more stringent condition than needed for magic state distillation.

\subsubsection{Code expansion}

This baseline is given by \cite{suzuki_q3de_2022}. The key idea of this method is to dynamically respond to a detected anomaly by increasing the code distance by $d_{\text{extra}}$ to a value at which the ray can be tolerated. The layout must therefore incorporate extra space around each surface code patch, which is available to use as routing space in the absence of anomalies but can be expanded into upon the detection of an anomaly.

If this extra added distance is large enough, we can in fact perform distillation in only $5 d_m$ cycles, as shown in Figure \ref{fig:baseline-expanded}, by routing between the logical qubits. When a ray hits, as shown in the bottom right image, a patch can expand, allowing the factory to continue distilling at the standard $6d_m$ cycle cost.

\begin{figure}[tp]
    \centering
    \includegraphics[width=\linewidth]{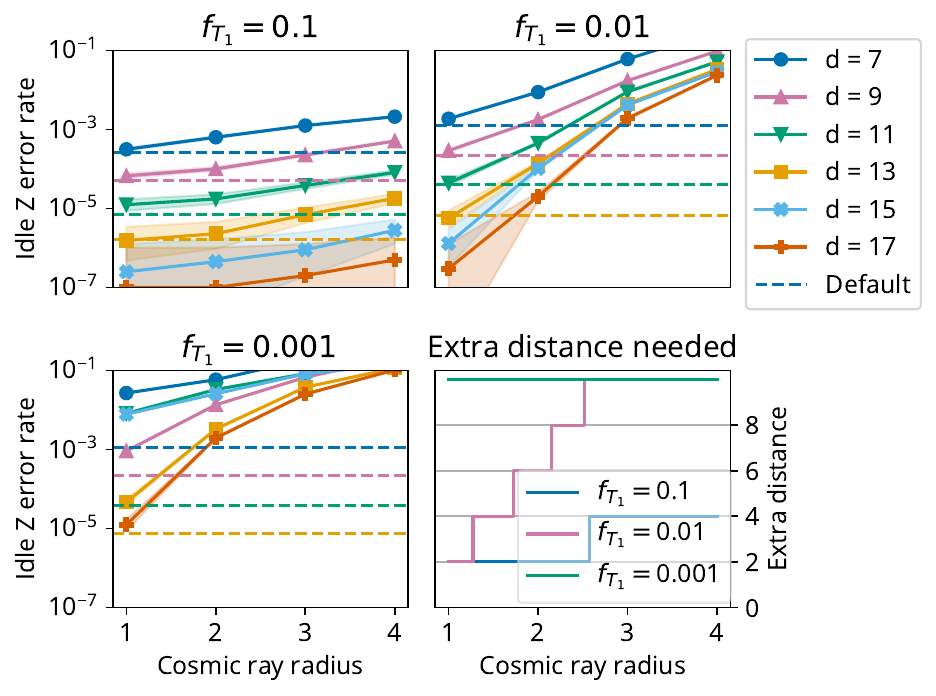}
    \caption{Extra code distance needed to tolerate \textbf{Direct} events of varying radius (assuming perfect decoder knowledge). Three plots show logical error rate as a function of ray radius, showing that larger rays degrade code distance more. Blue dashed line indicates $d=7$ error rate in the no-ray case. Each error rate plot shows the larger of either logical idle error rate (selected for $f_{T_1}=0.1$) or T rotation lattice surgery error rate (selected for $f_{T_1}=0.01$ and $0.001$), both obtained via Stim simulation. \emph{Bottom right:} We can extract the required additional code distance for a given ray radius and strength. This determines the overhead of the \emph{Code expansion} baseline.}
    \label{fig:expansion-required-distance}
\end{figure}

To determine the required $d_\text{extra}$ for the \textbf{Direct} model, we explicitly simulate the surface code idling logical error rate and the combined error rate associated with performing one of the T rotations via lattice surgery \cite{chamberland_universal_2022} (taking the worse error rate of the two) at various distances and with various-radius ray impacts. Results are shown in Figure \ref{fig:expansion-required-distance}. The distance required to tolerate rays increases very quickly as the radius increases. In our simulations we make the standard assumption that the decoder's knowledge of circuit-level error rates is correct; in a real-world scenario, performance may be worse \cite{hanks_decoding_2020, clader_impact_2021, berke_transmon_2022, iolius_performance_2022, carroll_subsystem_2024}. For the \textbf{Scrambling} model, we assume that the factory must be able to tolerate the worst-case scenario where a chain of qubits spanning the anomalous region are all disabled, so we use $d_\text{extra} = 2r_\text{CRE}$. We assume that $d_\text{extra}$ must be doubled if there is a significant chance of two simultaneous ray impacts on one patch, and so on for higher numbers of concurrent rays.

The related idea of quickly moving surface code patches away from anomaly locations \cite{sane_fight_2023} will have comparable space overhead, as it also requires enough available empty space to be able to move logical qubits away from an impact region.

\subsubsection{Distributed}

Our second comparison method is given by Ref. \cite{xu_distributed_2022}. The authors propose encoding surface code qubits in a higher-level erasure code using a distributed architecture (see Figure \ref{fig:baseline-distributed}). Several surface code qubits are linked together via distributed quantum connections, and checks are performed to encode these qubits into one higher-level logical qubit. A ray detection on one of the surface code qubits is treated as a heralded erasure error and the surface code qubit is reset. A code of distance $d$ can tolerate $d-1$ heralded erasure errors, assuming perfect heralding \cite{grassl_codes_1997}.

Ref. \cite{xu_distributed_2022} examines two code examples; an $[[n,k,d]] = [[4,1,2]]$ code (which uses 4 surface code qubits to encode 1 higher-level logical qubit with distance 2) which can correct at most 1 erasure, and a $[[7,1,3]]$ code that can correct at most 2 simultaneous erasures. These codes additionally require at least one \emph{ancilla} surface code qubit, bringing their total physical qubit overheads to 5x and 8x, respectively. In this work, we additionally consider $[[11,1,4]]$ and $[[17,1,5]]$ codes \cite{steane_simple_1996} when necessary to guarantee operation under higher rates of cosmic ray impacts. This method requires building an extensive distributed quantum architecture, adding an additional layer of complexity to device fabrication. 

\begin{figure}[tp]
    \centering
    \includegraphics[width=\linewidth]{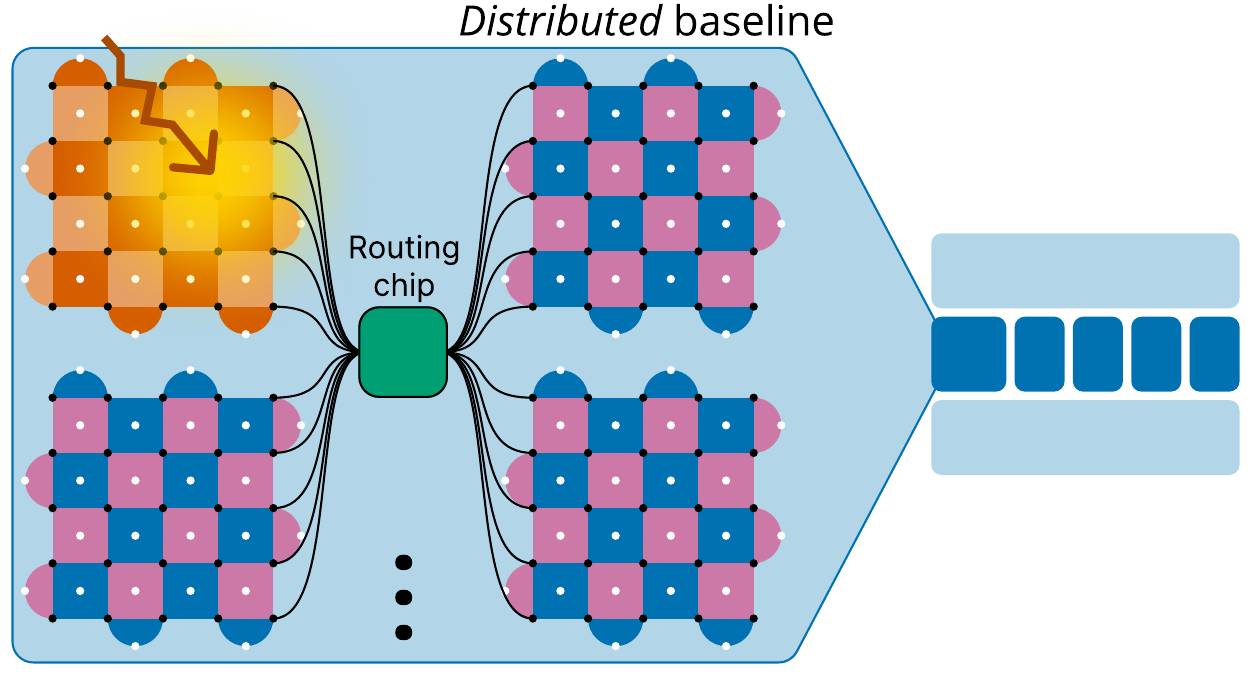}
    \caption{Schematic of the \emph{Distributed} baseline: each surface code patch is encoded in a higher-level code, which is triggered upon detection of a ray and can correct it by treating the anomaly as a heralded erasure error.}
    \label{fig:baseline-distributed}
\end{figure}

Ref. \cite{xu_distributed_2022} proposes keeping backup surface code chips in reserve to quickly be swapped in upon the detection and correction of an erasure, instead of needing to wait for the impacted chip to recover from the ray. This would significantly reduce $T_{\text{offline}}$ at the cost of more physical qubits. We do not explicitly study this scenario because all of the methods compared here could use this same strategy to reduce $T_{\text{offline}}$, so it would not significantly change comparative performance.

\subsection{Baselines require instant ray detection}\label{sec:evaluation/baselines-require-instant}

In Section \ref{sec:detecting} we found that, under our more realistic noise model and in the absence of hardware modifications, detecting ray impacts via syndrome counting is more difficult than assumed in prior works, especially if cosmic rays have decreased intensity or radius (which may occur as hardware mitigation strategies become more effective). 

In fact, weaker burst errors may counterintuitively be more challenging for the baseline methods, which require immediate and accurate detection of burst events.  The \emph{Code expansion} baseline relies on the ability to rewind the decoder to the moment of ray impact, but the authors note that the decoder cannot be rolled back beyond the most recent logical measurement operation \cite{suzuki_q3de_2022}; in magic state distillation, every operation involves a logical measurement, so this implies that detection must happen immediately or the ray cannot be mitigated. The \emph{Distributed} baseline relies on the assumption of immediate detection of any additional erasures that occur \emph{during} the recovery from the first erasure.

\begin{figure*}[hpt]
    \centering
    \includegraphics[width=\textwidth]{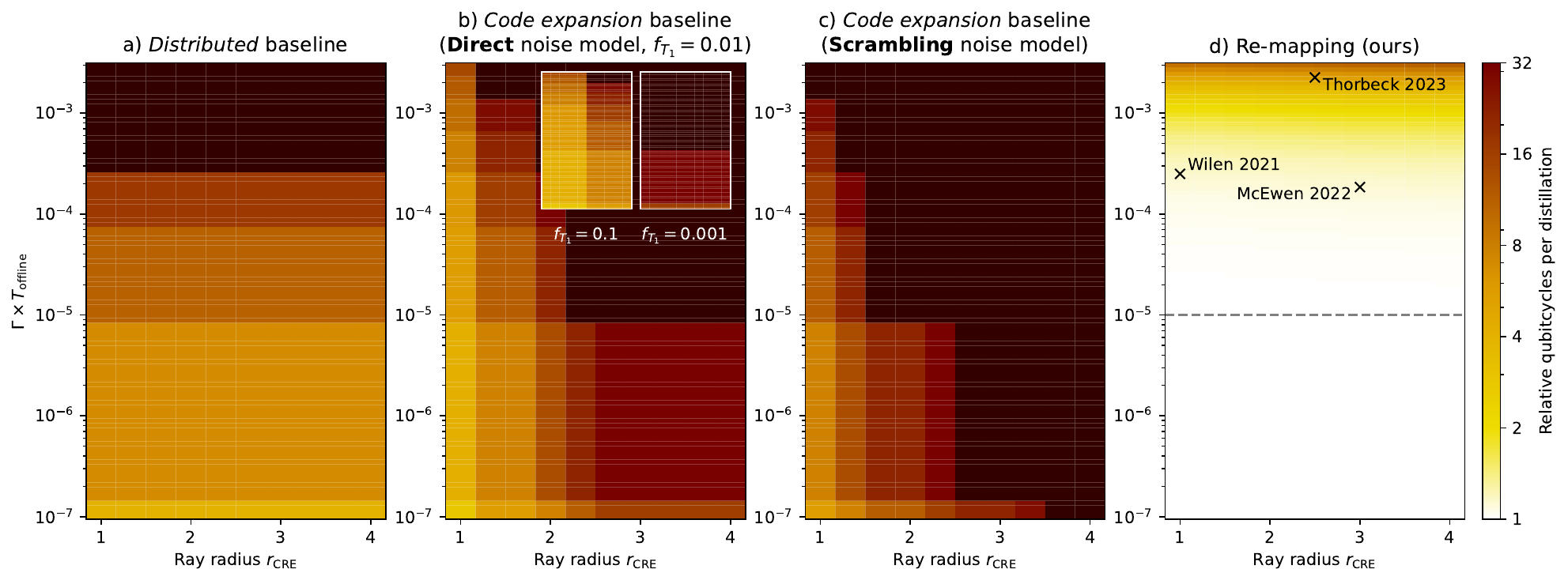}
    \caption{Quantifying the relative qubitcycle costs of cosmic ray mitigation methods in magic state factories, assuming ideal detection of ray impacts. Under ideal detection, the overheads of all methods are determined by the anomaly size (ray radius) and the value $\Gamma \times T_{\text{offline}},$ which determines the expected fraction of time that the factory is recovering from multiple ray impacts simultaneously. Our re-mapping method has significantly lower overhead across most of this parameter space. For low values of $\Gamma \times T_{\text{offline}},$ the re-mapping method incurs virtually no overhead, while for high values, the overhead is orders of magnitude lower than that of the baselines. Annotations on the rightmost plot mark the values of $\Gamma \times T_{\text{offline}}$ and $r_{\text{CRE}}$ observed in experiments \cite{wilen_correlated_2021, mcewen_resolving_2022, thorbeck_two-level-system_2023}. \textbf{Scrambling} model references \cite{wilen_correlated_2021} and \cite{thorbeck_two-level-system_2023} are shown assuming $T_{\text{offline}} = 1$ s. Gray dashed line incidates value of $\Gamma \times T_{\text{offline}}$ used in Figure \ref{fig:eval-realistic-detection}.}
    \label{fig:eval-ideal}
\end{figure*}

In contrast, our method is quite resilient to longer detection times, as the buffer for distilled states extends the maximum latency tolerable. Our method is philosophically different from the baseline methods: while the baseline methods aim to preserve the operation of all logical patches even under burst errors, we allow the burst errors to disrupt the patches temporarily and simply incur a relatively small time overhead upon a burst error event. The baselines, however, cannot guarantee the performance of all logical patches if burst errors are not immediately detected.


We thus compare our method to the two baselines under the generous assumption that a harmful ray impact can always be detected quickly enough for the method to be effectively implemented. This ideal detection could potentially be implemented by e.g. surrounding the chip with muon detectors \cite{mariani_mitigation_2023} at the cost of significantly increasing hardware complexity. 

\subsection{Calculating remapping overhead}

In this evaluation we consider the $(d_X, d_Z, d_m) = (7,3,3)$ distillation factory from \cite{litinski_magic_2019}, which is the smallest set of distances evaluated in that work. We note that our re-mapping method will perform comparably better with larger code distances, because any burst error event will lead to a relatively smaller area of the factory being turned offline, leading to \emph{decreasing} additional overheads from mitigation (while the baselines will have cost \emph{increasing} with the default footprint of the factory, as their overheads come from adding additional redundant space).

Most undetected ray impacts would be catastrophic to the program (Figure \ref{fig:expansion-required-distance}), so we must have a sufficiently reliable detection method. With this in mind, we therefore make the assumption that our windowed detector is tuned to always detect a ray in regions where it would harm the distillation (and may additionally trigger detection events in adjacent windows where the ray does not actually cause detrimental effects). In our evaluations, we thus do not quantify the \emph{fidelity} impact of a cosmic ray, but rather the increase in \emph{spacetime} overhead from turning off parts of the factory and re-mapping.

For a given cosmic ray impact location and a set of detector parameters (window size and threshold), we can calculate the chance that each logical patch is turned off (which happens if any detecting window in the patch is triggered). We then sample sets of offline patches from this distribution. For each, we find the optimal re-mapping the factory and obtain the cost (the number of cycles required per distillation). If a set of offline patches does not permit any distillation, we record this separately. Over many simulations, we obtain the probability distribution of distillation time overheads resulting from this cosmic ray, as well as the probability of the factory being inoperable for the duration of the ray.

We can repeat this process for many cosmic rays to determine the average inoperable chance and average operable overhead of a single cosmic ray impact anywhere on the chip.

\subsection{Qubitcycle cost as an evaluation metric}\label{sec:evalulation/qubitcycles}

We quantify distillation cost in terms of \emph{qubitcycles}, the product of the factory's physical qubit footprint and the number of stabilizer measurement rounds needed to distill one magic state. Spacetime volume is an appropriate metric because the baselines increase the space cost of a factory, while our method instead reduces the output rate of a factory. We now show that these can be thought of equivalently:

Depending on the desired rate of $T$ production, the program allocates space to $N_{\text{F}}$ magic state factories. Each of these factories produce magic states at rate $r_{\text{T}}$ and have a physical qubit footprint of $n_q$. Thus, the total T output rate of the distillation region is $N_{\text{F}}r_{\text{T}}$ with $n_q/r_{\text{T}}$ qubitcycles per T. In the presence of cosmic rays, the mitigation method can either increase $n_q$ (e.g., baselines) or decrease $r_{\text{T}}$ (e.g., our method). If the space cost \emph{increases} from $n_q \to f n_q$  we can maintain the original total output rate by still using $N_{\text{F}}$ factories, but now must use $f n_q N_{\text{F}}$ total physical qubits. The average qubitcycle cost is now $f n_q/r_{\text{T}}$. Similarly, if themitigation method  \emph{decreases} the output rate  $r_{\text{T}} \to \frac{1}{f}r_{\text{T}}$  we can now achieve the original total output rate by increasing the number of factories $N_{\text{F}} \to \frac{1}{f} N_{\text{F}}$. We thus use $f n_q N_{\text{F}}$ total physical qubits so that the average qubitcycle cost is again $f n_q/r_{\text{T}}$. 

Although the two methods have different impacts on a factory's space cost or production rate, they ultimately can be tuned to yield the exact same results from the program-level perspective. 
Thus, the ``qubitcycles per T state'' metric is a concise way to compare the costs of different methods.

\section{Results}\label{sec:results}

In this section, we examine distillation costs under cosmic ray events for the baselines and our method. We show that our method is vastly cheaper than baselines when burst errors can be detected perfectly. We then investigate performance of our method when detection latency is nonnegligible, where existing baselines would fail entirely.

\subsection{Baseline comparison under ideal detection}

As discussed in Section \ref{sec:evaluation/baselines-require-instant}, both baselines can only be used in magic state factories under the assumption of immediate and perfect detection of anomalous events. For our method, this means that we do not need any buffer space for distilled T states, as we can always simply stop the current distillation immediately upon a ray impact.

Results for the two baselines and our method are shown in Figure \ref{fig:eval-ideal}. In these plots, a cost of 1 corresponds to the same qubitcycle cost per distillation as the no-ray default. Because we assume that all rays are catastrophic, the overhead of the \emph{Distributed baseline} (a) only depends on $\Gamma \times T_{\text{offline}}$ and not on ray radius. The \emph{Code expansion} baseline also depends on the worst-case number of simultaneous rays, but also depends on the ray radius and the choice of noise model, as these determine the amount of extra code distance required (Figure \ref{fig:expansion-required-distance}). Finally, our method depends almost entirely on $\Gamma \times T_{\text{offline}}$, but there is a small dependence on the radius, as smaller rays take a smaller part of the factory offline. 

For all regions of the parameter space, our method vastly outperforms the two baselines. For the $f_{T_1} = 0.01$ \textbf{Direct} model corresponding to \cite{mcewen_resolving_2022}, we reduce the relative qubitcycle cost from $16.2$ to $1.1$ ($14.6\times$ improvement); across the entire parameter space for $f_{T_1} = 0.01$, we achieve qubitcycle reductions between $1.6\times$ and $87.3\times$ (geomean $11.1\times$ improvement) compared to the best baseline cost. For the weaker $f_{T_1} = 0.1$ \textbf{Direct} rays, we achieve qubitcycle reductions from $1.5\times$ to $18.1\times$ (geomean $6.5\times$). For the stronger $f_{T_1} = 0.001$ \textbf{Direct} rays, we achieve qubitcycle reductions from $4.0\times$ to $91.1\times$ (geomean $13.9\times$). For the \textbf{Scrambling} model, we achieve reductions for the model parameters in \cite{wilen_correlated_2021} and \cite{thorbeck_two-level-system_2023} of $5.3\times$ and $7.6\times$ respectively, and across \textbf{Scrambling} parameter space we improve upon the better baseline cost by $4.0\times$ to $205.5\times$ (geomean $13.8\times$).

\subsection{Costs under realistic detection}

We now examine the cost of our method in the context of realistic syndrome-measurement-based detection as discussed in Section \ref{sec:detecting}. As shown in the previous subsection, the time overhead of our method is quite small; the cost here is therefore primarily due to the physical qubit cost associated with maintaining enough buffer space for distilled T states to account for detection latency. For this analysis, we assume that each T state is stored in a distance 11 logical qubit.

For the \textbf{Direct} model, detection latency depends on both $f_{T_1}$ and $R_{\text{CRE}}$ (see Figure \ref{fig:detection-direct}). Relative qubitcycle cost results are shown in Figure \ref{fig:eval-realistic-detection} (top). For most of the studied parameter space, including that near the observed values from \cite{mcewen_resolving_2022}, our method has relatively low cost near the ideal value of $1\times$. However, as rays become smaller and weaker, detection becomes much more difficult, and the factory's qubitcycles increase dramatically as the required buffer space increases.

For the \textbf{Scrambling} model, we extract an approximate maximum detection latency from Figure \ref{fig:detection-scrambling} and use it to determine qubitcycle costs in our method. The results are shown in Figure \ref{fig:eval-realistic-detection} (bottom). Reliable detection is much more difficult under this noise model, leading to much higher costs over $100\times$ the default as ray radius decreases. 

While our method vastly outperforms the baselines in the ideal detection case, and works even when ray detection has non-negligible latency, it is clear that the overhead can still be prohibitive in the regime of large detection latency. This is an important area of improvement required to enable our method in the regimes of weak (but still damaging) cosmic rays.

\begin{figure}
    \centering
    \includegraphics[width=\linewidth]{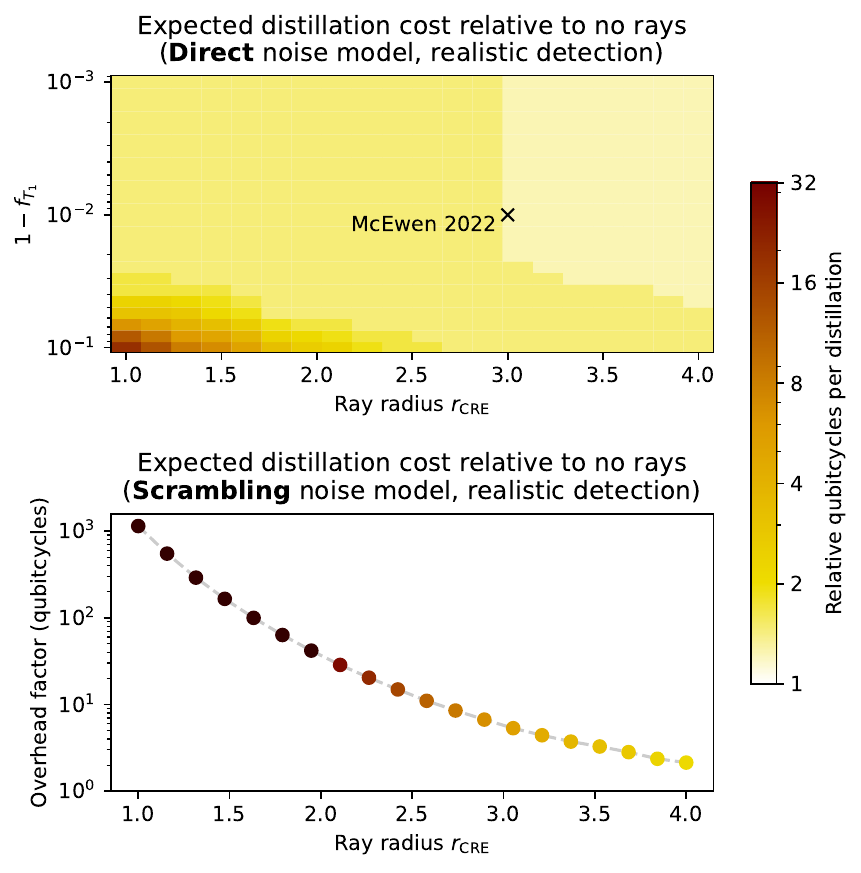}
    \caption{\emph{Top:} Distillation overhead of re-mapping method under varying detection latency for \textbf{Direct} model. Detection latencies are obtained from Figure \ref{fig:detection-direct}. Ray rate value $\Gamma \times T_{\text{offline}}$ is fixed at $10^{-5}$. \emph{Bottom:} Distillation overhead of re-mapping method under varying detection latency for \textbf{Scrambling} model. Detection latencies are obtained from Figure \ref{fig:detection-scrambling}.}
    \label{fig:eval-realistic-detection}
\end{figure}

\section{Conclusion}\label{sec:discussion}

In this work, we have demonstrated a much more efficient method to protect magic state factories from transient burst errors, reducing qubitcycle costs by orders of magnitude compared to preexisting baselines. As magic state factories are expected to take up to 95\% of the qubitcycles of future quantum programs \cite{beverland_assessing_2022, babbush_encoding_2018, blunt_compilation_2024}, this may be key to enabling practical quantum computation on superconducting hardware.

The idea of re-mapping magic state factories to avoid bad areas of the device is not specific to cosmic ray-induced errors. This approach could be used to allow for in-situ partial device recalibration due to other error sources as well, such as individual fluctuating TLSs or gate miscalibrations.

While we significantly reduced the overhead of cosmic ray mitigation in magic state factories, our method is not directly applicable to the program qubits that store logical information. Compared to previous work, we used more physically accurate noise models and highlighted the difficulty of quickly detecting a cosmic ray impact. We have effectively solved the problem of cosmic rays in magic state factories, but our work also shows that more research is needed to mitigate these events on the computational qubits.

In the future, shielding may be necessary to protect from the majority of disruptive burst events. Our method is still highly useful in the case of significantly reduced event rate, because it incurs no extra time overhead except when a ray hits, meaning that there is no downside to using it when there is any nonzero chance of a burst event.

As quantum computing approaches the first true fault-tolerant devices, tailoring systems design to specific subroutines will become increasingly critical to reduce the intimidating overheads of quantum error correction. Our work demonstrates that significant gains can be made by considering the unique characteristics of fault-tolerant program subroutines.

\section*{Contributions}

CK proposed the problem of time-varying noise in magic state factories. JDC developed the mitigation method for cosmic rays, created the noise models, and wrote evaluation code. JV wrote the lattice surgery Stim simulation code. JV and SFL contributed to earlier simulation code. All authors contributed to guiding the project and revising the manuscript.

\section*{Acknowledgements}

We thank Jonathan Baker for helpful discussions on factory re-mapping, and Matt McEwen for an insightful chat on cosmic rays and gap engineering.

\section*{Data availability}

The code to perform all experiments and generate figures is publicly available at \href{https://github.com/jasonchadwick/ray-delay}{github.com/jasonchadwick/ray-delay}.

\bibliography{references.bib}


\end{document}